\documentclass[pra,aps,amsmath,reprint,10pt]{revtex4-2}
\usepackage{natbib}
\usepackage{graphicx}
\usepackage{xcolor}
\usepackage{amsfonts}
\usepackage{amsmath}
\usepackage{hyperref}
\usepackage{amssymb}
\usepackage{amsthm}
\usepackage{bbold}

\usepackage{braket}
\usepackage{tikz}
\usetikzlibrary{positioning}
\usetikzlibrary{shapes,arrows,arrows,positioning,fit}

\begin{document} 
\title{Fractional Conformal Map, Qubit Dynamics and the Leggett-Garg Inequality }

% If your conference documentclass or package defines these macros,
% change these macros to different names.
\newcommand*{\affaddr}[1]{#1} % No op here. Customize it for different styles.
\newcommand*{\affmark}[1][*]{\textsuperscript{#1}}

\author{Sourav Paul \affmark[1] }
\email{sp20rs034@iiserkol.ac.in; souravpl2012@gmail.com}

\author{Anant Vijay Varma \affmark[1,2]}
\email{anantvijay.cct@gmail.com}

\author{Sourin Das \affmark[1]}
\email{sourin@iiserkol.ac.in\\}

\affiliation{
\affaddr{\affmark[1] Indian Institute of Science Education and Research Kolkata, Mohanpur, Nadia 741246, West Bengal, India.}\\
\affaddr{\affmark[2] Department of Chemistry, Ben-Gurion University of the Negev, Beer-Sheva 84105, Israel}}

\begin{abstract}
Any pure state of a qubit can be geometrically represented as a point on the extended complex plane through stereographic projection. By employing successive conformal maps on the extended complex plane, we can generate an effective discrete-time evolution of the pure states of the qubit. This work focuses on a subset of analytic maps known as fractional linear conformal maps. We show that these maps serve as a unifying framework for a diverse range of quantum-inspired conceivable dynamics, including (i)  unitary dynamics,(ii) non-unitary but linear dynamics and (iii) \ non-unitary and non-linear dynamics where linearity (non-linearity) refers to the action of the discrete time evolution operator on the Hilbert space. We provide a characterization of these maps in terms of Leggett-Garg Inequality complemented with No-signaling in Time (NSIT) and Arrow of Time (AoT) conditions. 
\end{abstract}
\maketitle

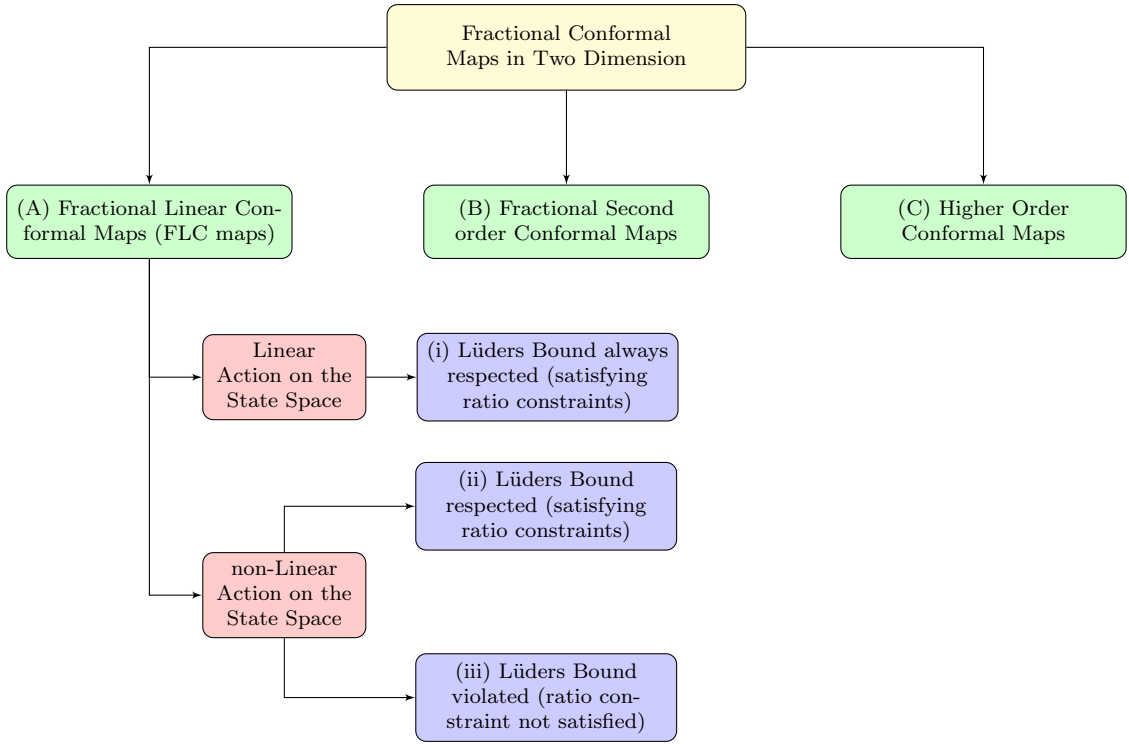
\begin{figure*}
 	\tikzstyle{decision} = [diamond, draw, fill=blue!20, 
 	text width=5.5em, text badly centered, inner sep=0pt]
 	
  \tikzstyle{block} = [rectangle, draw, fill=blue!20, 
 	text width=10em, text centered, rounded corners, minimum height=2em]
 	
  \tikzstyle{line} = [draw, -latex']
 	
  \tikzstyle{cloud} = [draw, ellipse,fill=red!20,
 	minimum height=4em]
 	
  \tikzstyle{decisionn} = [diamond, draw, fill=red!20, 
 	text width=4em, text badly centered, node distance=4cm, inner sep=0pt]
 	
  \tikzstyle{blockk} = [rectangle, draw, fill=green!20, 
 	text width=11em, text centered, rounded corners, minimum height=3em]
 	
  \tikzstyle{blockkk} = [rectangle, draw, fill=red!20, 
 	text width=6em, text centered, rounded corners, minimum height=2em]

  \tikzstyle{blockkkk} = [rectangle, draw, fill=yellow!20, 
 	text width=14em, text centered, rounded corners, minimum height=3.5em]

%%%%%%%%%%%%%%%%%%%%%%%%%%%%%%%%%%%%%%%%%%%%%%%%%%

\begin{center}
 	\begin{tikzpicture}[node distance = 1.75cm,  auto, >=stealth, every node/.style={ auto, font=\footnotesize, anchor=center, >=stealth}]\label{ams1}
  
 	%%%%%%%%% Place nodes

 	\node [blockkkk] (init) {Fractional Conformal Maps in Two Dimension};

   	\node [blockk, below left=1.25cm and 1.25cm of init] (A) {(A) Fractional Linear Conformal Maps (FLC maps)};
    
 	\node [blockk, below =1.25cm of init] (B) {(B) Fractional Second order Conformal Maps};

 	\node [blockk, below right=1.25cm and 1.25cm of init] (C) {(C) Higher Order Conformal Maps};

 	\path [line] (init) -| (A);
 	\path [line] (init) -- (B);
        \path [line] (init) -| (C);

%%%%%%%%%%%%%%%%%%%%%%%%%%%%%%%%%%%%%%%%%%%%%

	\node [blockkk, below right = 1.0cm and -1.2cm of A] (Linear) {Linear Action on the State Space};
 
    \path [line] (A) |- (Linear);

    \node [blockkk, below right = 3.9cm and -1.2cm of A] (nonLinear) {non-Linear Action on the State Space};

    \path [line] (A) |- (nonLinear);

%%%%%%%%%%%%%%%%%%%%%%%%%%%%%%%%%%%%%%%%%

   %\node [blockkk, right of =B, node distance=4.5cm] (Chaos) {Ambiguous discrimination of the dynamics owing to chaos};
  
    %\path [line] (B) -- (Chaos);

%%%%%%%%%%%%%%%%%%%%%%%%%%%%%%%%%%%%%%%%%%%%%%%%%
    \node [block, right of =Linear, node distance=3.5cm] (Luder) {(i) L\"{u}ders Bound always respected (satisfying ratio constraints)};

    \path [line] (Linear) -- (Luder);

%%%%%%%%%%%%%%%%%%%%%%%%%%%%%%%%%%%%%%%%%%%%%%%%%%%
    \node [block, above right = 0.02cm and 0.65cm of nonLinear] (Luder1) {(ii) L\"{u}ders Bound respected (satisfying ratio constraints)};
    
     \path [line] (nonLinear) |- (Luder1);
    
 	\node [block, below right = 0.2cm and 0.65cm of nonLinear] (Luder2) {(iii) L\"{u}ders Bound violated (ratio constraint not satisfied)};

    \path [line] (nonLinear) |- (Luder2);

 	\end{tikzpicture}
  \end{center}
 	\caption{Fractional Conformal Maps in two dimension and the L\"{u}ders bound}
  \label{fig1}
 \end{figure*}

%----------------------------------------------------------------------
\section{Introduction}{\label{I}}
%----------------------------------------------------------------------
It is well known that the Bloch sphere can be identified with the Riemann sphere. The Riemann sphere, also known as the extended complex plane, serves as a one-point compactification of the complex plane (denoted as $C$) \cite{najarbashi,Nakahara2008, Francis1983,Elias2003}. This compactification is represented as  $C \  \cup  \  \infty = \tilde{C}$. Through stereographic projection, any point on the Bloch sphere can be mapped to a complex number  $z$  on the extended complex plane. This geometric representation has been extensively explored in various contexts, as highlighted in works like \cite{KimLee,najarbashi,Gilyen, KissT}. The successive application of a conformal map to a point on the extended complex plane, can be viewed as a discrete-time evolution of the Bloch vector on the Bloch sphere, representing the pure state. The complete set of conformal maps on the complex plane is characterized by locally invertible complex analytic functions, commonly known as M\"{o}bius transformations \cite{Nevanlinna, Ahlfors, Rudin, Yau2016}.
\par The connection between the unitary evolution of a qubit, represented by a Bloch vector in two dimensions, and fractional linear conformal maps has been previously explored by Kim and Lee \cite{KimLee}. In this study, we extend this approach by delving into the exploration of all conceivable ``fractional linear conformal maps" (FLC maps) in two dimensions and their classification based on the temporal correlations induced by such discrete-time qubit dynamics. 
\par To probe the quantum nature of discrete-time qubit dynamics via temporal correlations, we employ the Leggett-Garg Inequality (LGI) \cite{LG1985, Leggett2002, Leggett2008}. In a simplified scenario involving three projective measurements and two steps of evolutions, we define the LGI using the LG parameter $K_{3}$ as: $-3 \leq K_{3} = C_{12} + C_{23} - C_{13} \leq 1$, where $C_{ij}$ denotes two-time correlations. The violation of $K_3$ parameter serves as an indicator of non-classical behavior along with the conditions of No-signaling in Time (NSIT) and Arrow of Time (AoT) \cite{EmaryC, Kofler}, with the upper bound $3/2$ of $K_3$ referred to as the L\"{u}ders bound. Notably, recent theoretical studies and experimental observations have explored the potential violation of the L{\"{u}}ders bound, particularly in non-Hermitian systems \cite{Tusun, Zhan, PLu2023, quinn2023, Anant, Yogesh,Usha, Pan}.
\par We categorize the parameter space of these FLC maps based on (a) the temporal correlations arising from discrete-time evolution and (b) the Linear  or non-Linear action on the Hilbert space. (\textcolor{blue}{see {{FIG. \ref{fig1}}}}). Employing a discrete-time version of LGI, we delineate the entire parameter space of these maps into three distinct classes:
(i) exhibiting linear action on the Hilbert space  respecting the L{\"{u}}ders bound,
(ii) exhibiting non-linear action on the Hilbert space satisfying the L{\"{u}}ders bound,
(iii) exhibiting non-linear action on the Hilbert space violating the L{\"{u}}ders bound.
\par The article is organized as follows: In section \ref{II}, we introduce FLC maps, discussing the relationship between the extended complex plane and the Bloch sphere through stereographic projection. This section also comprehensively presents the dynamics of the pure state of the qubit induced by FLC maps. Moving on to section \ref{III}, we devote our discussion to the interplay of linear and non-linear actions on the pure qubit state induced by these maps. In section \ref{IV} we introduce the formalism for calculating the LGI for FLC maps and present analytical results. Finally, section \ref{V} is reserved for discussions and concluding remarks.

%\par \textcolor{blue}{The remainder of this article is organized as follows. In Sec.~\ref{LCM}, we discuss the linear conformal maps. After that in Sec.~\ref{secII} we discuss the Bloch Sphere states and its correspondence to the extended complex plane via Stereographic projection and how evolution of states in Bloch sphere can be thought as an iterative actions of such maps in the extended complex plane. Sec.~\ref{secIII} is devoted to discrete time LGI analysis for above-mentioned maps. In  Sec.~\ref{secIV} we present our results. Sec.~\ref{secV} is dedicated to discussion and conclsion.
%}

%\begin{figure}[htb!]
%    \centering
%\includegraphics[width=3.9cm,height=3.5cm]{julia_set.pdf} 
%\includegraphics[width=3.9cm,height=3.5cm]{julia_set_1.pdf}  \hspace{0.75em} 
%\caption{\textbf{Julia Set and Fatou Set for one-variable map plotted in complex plane.} Julia Set(Blue) consisting initial sensitive points, Fatou set (white) consisting of regular points. Fig (a) and (b) correspond to the complex map $f(z) = \frac{z^2 + \imath \tan 0.4}{\imath z^2 \tan 0.4 + 1}$ and $f(z) = \frac{z^2 + \imath \tan 0.55}{\imath z^2 \tan 0.55 + 1}$ respectively.}
%\label{fig1}
%\end{figure}
%%%%%%%%%%%%%%%%%%%%%%%%%%%%%%%%%%%%%%%%%%%%%%%%%%%%%%%%%%%%%%%%%%%%%%%%%%%%%%%%%%%%%%%%%5

%\section{Fractional linear conformal map on a complex plane} 
\section{FLC maps induced dynamics\label{II}}
The stereographic projection is established by identifying the Bloch sphere with the Riemann sphere, enabling the definition of a projection $S:\mathcal{H} \longrightarrow \tilde{C} $ from the two-dimensional projective Hilbert space $\mathcal{H}$  to the extended complex plane $\tilde{C}$ \cite{KimLee}.
%It is known that the Riemann sphere, also called the extended complex
%plane, is a 1-point compactification of the complex plane (denoted by C)~ \cite{KimLee},
%$C \  \cup  \  \infty = \tilde{C}$. %The Stereographic projection maps any point on the Bloch sphere to a complex number  $z$ on the extended complex plane. By identifying the Bloch sphere with the
%Riemann sphere, one can define a projection
%$S:\mathcal{H} \longrightarrow \tilde{C} $ 
%from the two-dimensional projective Hilbert space $\mathcal{H}$ to the extended complex plane $\tilde{C}$ ~ \cite{KimLee}.
%Conformal maps whose action is linear on the complex plane represented by $z$ can be written as:
A mathematical map, denoted as
\begin{equation}
f(z) = \frac{a z + b}{c z + d},
\label{E00}
\end{equation}
where $a$, $b$, $c$, and $d$ are complex numbers with $ad-bc \not =0 $, is characterized as a `fractional linear conformal map (FLC map)' %`Mobi\"{u}s Transform' or `Bilinear Transform'
\cite{Rudin, Ahlfors}. In the Bloch sphere, the pure state of a qubit is represented by
$\ket{\psi} = (\zeta_1, \zeta_2) =  N (z,1)^{T}$, where $N=1/\sqrt{\vert z \vert^2 + 1}$ along with state-correspondent point on the extended complex plane being $z ( = \zeta_1/\zeta_2$) \cite{KimLee},neglecting the overall phase. The discrete-time evolved state, corresponding to $z  \mapsto f(z)$, is expressed as 

\begin{equation}
\ket{\psi^{'}} = \frac{1}{\sqrt{\vert f(z) \vert^2 + 1}} \  (f(z),1)^{T}
\label{E01}
\end{equation}
Equivalently, the operation induced by the fractional linear conformal map on the qubit state is represented as
\begin{equation}
\vert \psi^{'} \rangle = N_1 \Big( M \vert \psi \rangle \Big) = N_1 \begin{pmatrix}
a & b \\ c & d
\end{pmatrix}  \vert \psi \rangle 
= N_1 \begin{pmatrix}
a z + b \\
c z + d
\end{pmatrix}
\label{E02}
\end{equation}
where $M = \begin{pmatrix}
    a & b\\
    c & d
\end{pmatrix}$ 
is the matrix representation corresponding to the FLC map $f(z)$ with $N_1$ being the overall normalization of the qubit after the operation. This transformation ensures at least Positivity and Trace Preserving (PTP) properties, preserving the Hermiticity of the corresponding qubit state density matrix \cite{Lindblad,Choi}.
%Therefore, the maps in eqn. (\ref{E00}) defines full set of transformations/maps we have considered in this paper. \textcolor{red}{In what follows we divide the maps based on the allowed values of the complex numbers $a,b,c,$ and $d$, into different sub-sets, which correspond to different physical dynamics. \textbf{I have a doubt in this line whether it is relevant to write it here?}}
The following relation schematically shows the direct correspondence between the Bloch sphere state and a point on the extended complex plane.
\begin{equation}
\ket{\psi} \longleftrightarrow z \mapsto z' = f(z) \longleftrightarrow \ket{\psi}'
\label{E04}
\end{equation}

%In order to do that we first divide the maps $f(z)$ into two sets. \\ \\
%(A) \ \  $  \{ f(z): $  the action of map $f$ on state $z$ is 'linear' at the level of Hilbert space. $ \} $  \\  \\
%(B) \ \  $  \{ f(z):$  the action of map $f$ on state $z$ is 'non-linear' at the level of Hilbert space. $ \} $  \\

%he action of the map $f(z)$ is linear at the level of the Hilbert space if

%\begin{equation}
%d = a^{*}, \quad c = -b^{*},
%\label{E02}
%\end{equation}

%together with the constraint $|a|^2 + |b|^2 = 1$.

%(\textcolor{red}{ I think we can explicitly show here for which range of the complex numbers a,b,c,d we have the linear action of the state space level. btw what is correct terminology here'at the level of Hilbert space' or 'on the state space'. plse do some digging.})

%(\textcolor{red}{ defining the set of linear operations at the level of the Hilbert spce, would clear the defintion what we mean by 'linear'}
%%%%%%%%%%%%%%%%%%%%%%%%%%%%%%%%%%%%%%%%%%%%%%%%%%%%%%%%%%%%%%%%%%%%%%%%%%%%%%%%%%%%%%%%%%%%%%%%%%%%%%%%%%%%%%%%%%%%%%%%%%%%%%%%%%%%%%%%%%%%%%%%%%%%%%%%%%%%%%%%%%%%%%%%%%

%\section{Linear(unitary) and Non-linear(non-unitary) action on the pure state qubit induced by fractional linear conformal map}
\section{Linear vs. Non-linear Actions on the hilbert space \label{III}}

Action of a linear operator $\hat{O}$ on a pure state $\vert \psi \rangle = \eta_1 \vert \psi_1 \rangle + \eta_2 \vert \psi_2 \rangle$ can be expressed as 
\begin{equation}
\hat{O}\vert \psi \rangle = \eta_1 \ \hat{O} \vert \psi_1 \rangle + \eta_2 \ \hat{O} \vert \psi_2 \rangle ,
\label{E05}
\end{equation}
where $\vert \psi \rangle$ is decomposed in the orthonormal basis  \{ $\vert \psi_1 \rangle$ , $\vert \psi_2 \rangle$\} with $\vert \eta_1 \vert^2 + \vert \eta_2 \vert^2 = 1$. 
%Notably the states $U\vert \psi \rangle$, $U\vert \psi_1 \rangle$ and $U\vert \psi_2 \rangle$ must be pure states or transformable into pure states for stereographic projection to be applicable. 
A class of linear operator acting on the two dimensional Hilbert space can be written  as 
\begin{equation}
 \hat{O} = \begin{pmatrix}
a & b\\
-b^{*} & a^{*}
\end{pmatrix} 
\label{E051}
\end{equation} with $\vert a \vert^2 + \vert b \vert^2 = r$, where $a,b \in \mathbb{C}, r\in \mathbb{R}$. For determinant $r=1$, $\hat{O}$ is the \textit{unitary} operator, while for $r \neq 1$, $\hat{O}$ corresponds to an operator which is unitary $\circ$ scaling. We will show that the parameter space of FLC maps (defined in Eq.(\ref{E00})) includes both $r=1$ (unitary case explored by \cite{KimLee})and $r\neq 1$ case.  On the contrary, $\hat{O}$ can violate linearity given in Eq.(\ref{E05}) if one relaxes the constraint $\vert a \vert^2 + \vert b \vert^2 = r$  such that,
%\par A non-linear operator $\tilde{U} $ violates Eq.(\ref{E05}) for a pure state vector $\vert \psi \rangle = \eta_1 \vert \psi_1 \rangle + \eta_2 \vert \psi_2 \rangle$, leading to the  inequality
\begin{equation}
\frac{\hat{O}\vert \psi \rangle}{\sqrt{\langle \psi \vert \hat{O}^{\dagger} \hat{O} \vert \psi \rangle}} \neq \tilde{N} \Big(\eta_1 \frac{\hat{O}\vert \psi_1 \rangle}{\sqrt{\langle \psi_1 \vert \hat{O}^{\dagger} \hat{O} \vert \psi_1 \rangle}} + \eta_2 \frac{\hat{O}\vert \psi_2 \rangle}{\sqrt{\langle \psi_2 \vert \hat{O}^{\dagger} \hat{O} \vert \psi_2 \rangle}}\Big)
\label{E06}
\end{equation}
where $\tilde{N} = \Big\vert \Big\vert \eta_1 \frac{\hat{O}\vert \psi_1 \rangle}{\sqrt{\langle \psi_1 \vert \hat{O}^{\dagger} \hat{O} \vert \psi_1 \rangle}} + \eta_2 \frac{\hat{O}\vert \psi_2 \rangle}{\sqrt{\langle \psi_2 \vert \hat{O}^{\dagger} \hat{O} \vert \psi_2 \rangle}} \Big\vert \Big\vert$. It is also shown that the parameter space of FLC maps also includes $\vert a \vert^2 + \vert b \vert^2 \neq r$ case. Although there can be a general class of non-linear operators not having the form of $\hat{O}$ but still following condition (\ref{E06}). 
\par In general, any quantum evolution must respect linearity though unitarity can be compromised for an open system. On the other hand, quantum evolution interrupted by quantum measurement, followed by post selection can give rise to an effective dynamics having non-linear action on the space of states {\cite{Anant, Yogesh}}. In the next section we establish that FLC maps cover all  possible quantum dynamics i.e. unitary, non-unitary but linear, non-unitary and non-linear.
\begin{figure*}[t]
    \centering
    \includegraphics[scale=0.45]{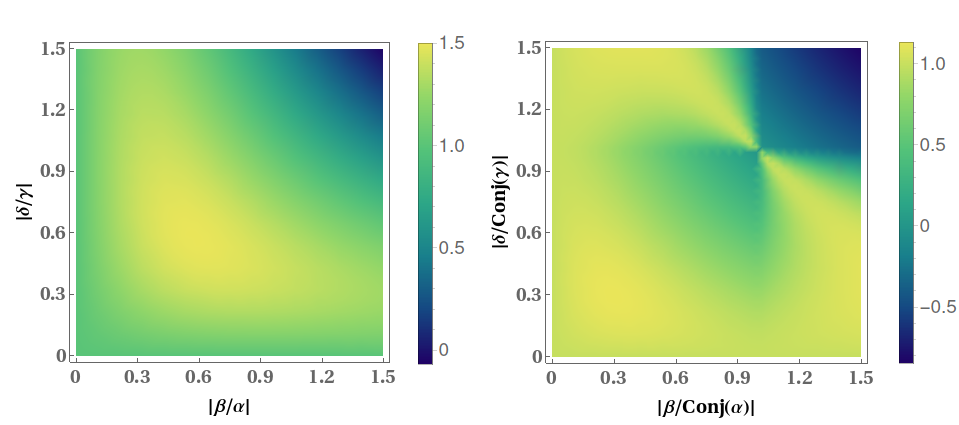}
    \caption{ Optimal LG $K_3$ values for a qubit undergoing dynamics induced by discrete FLC maps $f_{12}$, $f_{23}$, and $f_{13}$ (refer to \textcolor{blue}{Appendix-\ref{AppC2}}). (a) Left diagram corresponds to the maps $f_{12}(z) = \frac{\alpha z + \beta}{\beta z + \alpha}$ and $f_{23}(z) = \frac{\gamma z + \delta}{\delta z + \gamma}$. (b) Right diagram corresponds to the maps $f_{12}(z) = \frac{\alpha z + \beta}{\beta^{*} z + \alpha^{*}}$ and $f_{23}(z) = \frac{\gamma z + \delta}{\delta^{*} z + \gamma^{*} }$. In both cases, the plot is independent of the initial qubit state on the Bloch sphere.
    }
    \label{fig2}
\end{figure*}
%%%%%%%%%%%%%%%%%%%%%%%%%%%%%%%%%%%%%%%%%%%%%%%%%%%%%%%%%%%%%%%%%%%%%%%%%%%%%%%%%%%%%%%%
\section{Discrete Time LGI with FLC Maps\label{IV}}
% \textcolor{red}{cite the paper and see the paper by (\textit{Clemente and Kofler)}}\\ \\
In this section we consider evaluating three time LG parameter $K_3$ for discrete time evolution induced by FLC maps , corresponding to the dichotomic measurement operator $\hat{Q}= \sigma_z$. Steps followed for inducing the discrete evolution is given below.\\ \\
\textit{\underline{Steps:}}
(\textit{i}) The state represented by $z=z_1$ on the extended complex plane is acted upon using the map $f_{12}(z)$. This map should be understood as the evolution of the system starting from time $t_1$ to time $t_2$.
\par (\textit{ii}) The subsequent effective evolution over the next time interval from $t_2$ to  $t_3$ is induced by the map $f_{23}(z)$ acting on the the state represented by $z = z_2 = f_{12}(z_1)$ at 
$t_2$.
\par (\textit{iii}) The composite evolution from $t_{1}$ to $t_{3}$ is induced by the composition of the above maps i.e. $f_{23} \circ f_{12}$ starting from the state at $t_{1}$. 
\begin{equation*}
z_2 = f_{12}(z_{1}) = \frac{a_{12} \ z_{1} + b_{12}}{c_{12} \ z_{1} + d_{12}}; \ \  z_3 = f_{23}(z_{2}) = \frac{a_{23} \ z_{2} + b_{23}}{c_{23} \ z_{2} + d_{23}}
\end{equation*}
\begin{equation}
z_3 = f_{23}(f_{12}(z_1)) = \frac{a_{13}z_1 + b_{13}}{c_{13} z_1 + d_{13}}
\label{E08}
\end{equation}
where, $a_{13} = a_{12}a_{23} + c_{12} b_{23}, \ b_{13} = b_{12}a_{23} + d_{12}b_{23}, \ c_{13} = a_{12}c_{23} + c_{12}d_{23}, \ d_{13} =  b_{12}c_{23} + d_{12}d_{23}$.
\subsection{Joint Probabilities ($P_{ij}$) and correlation functions ($C_{ij}$) \label{IV-A}}
In this subsection we define  the temporal correlations $C_{ij}$'s expressed in terms of joint probabilities $P_{ij}$'s which are required to evaluate the LG parameter for the dichotomic observable $\hat{Q} = \sigma_z$.
\begin{equation}
C_{ij} = \sum_{\hat{Q}(t_i/t_j) = \pm 1} \hat{Q}(t_i) \hat{Q}(t_j) P_{ij} ( \hat{Q}(t_i), \hat{Q}(t_j)) 
\label{E09}
\end{equation}
\begin{equation*}
\text{with} \quad  P_{ij} (\hat{Q}(t_i),\hat{Q}(t_j)) =    
\end{equation*}
\begin{equation}
\frac{(\vert \langle \pm_Q \vert e^{-i H (t_j-t_i)}\vert \pm_Q \rangle \vert^2) (\vert \langle \pm_Q \vert e^{-i H (t_i-t_1)}\vert \psi^{(0)} \rangle \vert^2 )}{\langle \pm_Q \vert e^{i H^{\dagger} (t_j-t_i)} e^{-i H (t_j-t_i)} \vert \pm_Q \rangle  \langle \pm_Q \vert e^{i H^{\dagger} (t_i-t_1)} e^{-i H (t_i-t_1)} \vert \psi^{(0)} \rangle}
\label{E10}
\end{equation}
where, $i<j \ \{i,j = 1,2,3\}$ and $\ket{\psi^{(0)}}$ is the initial state. Here $\hat{Q}(t_k)$ denotes the measurement outcome either $+1$ (corresponding to the $\ket{\uparrow} = (1, 0)^T$) or $-1$ (corresponding to the $\ket{\downarrow} = (0, 1)^T$)  of dichotomic observable $\hat{Q}  = \sigma_z$. We must point out that for our calculation, the state $e^{-i H (t_j-t_i)} \vert \phi \rangle$ is equivalent to the state $(f_{ij}(z_{\vert \phi \rangle}),1)^{T}$ where $z_{\vert \phi \rangle}$ is the point on the complex plane corresponding to the state $\vert \phi \rangle$ on the Bloch sphere via stereographic projection. Also $f_{ii}(z_{\vert \phi \rangle}) = z_{\vert \phi \rangle}$ for all $\{i = 1,2,3\}$.
\par \textit{\underline{Calculation of $C_{12}$} }: \ Assuming $t_1 = 0$, the initial state $\vert \psi^{(0)} \rangle( = \vert \psi(t_1=0)  \rangle)$ is expressed in the eigenbasis of $\hat{Q}$ as,
\begin{equation}
\ket{\psi^{(0)}} = \frac{1}{\sqrt{\vert z_{\ket{\psi^{(0)}}} \vert^2 + 1}} \ (z_{\ket{\psi^{(0)}}},1)^{T}.
\label{E11}
\end{equation}
The probabilities of obtaining the $\ket{\uparrow}$ or $\ket{\downarrow}$  eigenstates of the  measurement operator $\hat{Q} = \sigma_z$ at $t_1$ are then calculated (assuming $z_{\ket{\psi^{(0)}}}=re^{i \phi} $) as:

\begin{equation}
P_{t_{1}}(+) = |\braket{\uparrow|\psi^{(0)}}|^{2}= \frac{r^{2}}{1+r^{2}} 
\label{E12}
\end{equation}
\begin{equation}
P_{t_{1}}(-) = |\braket{\downarrow|\psi^{(0)}}|^{2}= \frac{1}{1+r^{2}}
\label{E13}
\end{equation}
Thereafter the two eigenstates (\{$\ket{\uparrow}$, $\ket{\downarrow} $\}) are evolved using the map 
$f_{12}(z)$, where $z = \infty$ for $\ket{\uparrow}$ and $z = 0$ for $\ket{\downarrow}$. This results in two states at time $t_2$ (denoted as $\ket{\psi(t_{2})}$) expressed as:
\begin{equation}
\ket{\psi(t_{2})}_{\ket{\uparrow}} = \frac{1}{\sqrt{\vert f(z=\infty) \vert^2 + 1}} \  (f(z=\infty),1)^{T} 
\label{E14}
\end{equation}
\begin{equation}
\ket{\psi(t_{2})}_{\ket{\downarrow}} =\frac{1}{\sqrt{\vert f(z=0) \vert^2 + 1}} \  (f(z=0),1)^{T},
\label{E15}
\end{equation}
The above equations can be re-written using eq.(\ref{E08}) as:
\begin{equation}
\ket{\psi(t_{2})}_{\ket{\uparrow}} = \frac{1}{\sqrt{\vert a_{12}/c_{12} \vert^2 + 1}} \  (a_{12}/c_{12},1)^{T}
\label{E16}
\end{equation}
\begin{equation}
\ket{\psi(t_{2})}_{\ket{\downarrow}} =\frac{1}{\sqrt{\vert b_{12}/d_{12} \vert^2 + 1}} \  (b_{12}/d_{12},1)^{T}.
\label{E17}
\end{equation}
Hence the joint probabilities can be evaluated as:
\begin{equation}
 P_{12}(+,+)= \frac{r^{2}}{1+r^{2}} \times 
\frac{|a_{12}/c_{12}|^{2}}{|a_{12}/c_{12}|^{2}+1},
 \label{E18}
\end{equation}

\begin{equation}
 P_{12}(+,-)=  \frac{r^{2}}{1+r^{2}} \times \frac{1}{|a_{12}/c_{12}|^{2}+1},
 \label{E19}
\end{equation}

\begin{equation}
 P_{12}(-,+)= \frac{1}{1+r^{2}} \times \frac{|b_{12}/d_{12}|^{2}}{|b_{12}/d_{12}|^{2}+1},
 \label{E20}
\end{equation}

\begin{equation}
 P_{12}(-,-)= \frac{1}{1+r^{2}} \times \frac{1}{|b_{12}/d_{12}|^{2}+1},
\label{E21}
\end{equation}
such that,
\begin{equation}
C_{12} = P_{12}(+.+) + P_{12}(\-,-)
-  P_{12}(-,+) - P_{12}(+,-)
\label{E22}
\end{equation}
which reduces to.
\begin{equation}
C_{12} = 1-2 z_{12} + 2 x_{12} (y_{12}+z_{12}-1),
\label{E23}
\end{equation}
where the  $x_{12},y_{12},z_{12}$ lie between 0 and 1. The explicit forms are $x_{12} = \frac{r^{2}}{1+r^{2}}$, $y_{12}= \frac{|a_{12}/c_{12}|^{2}}{|a_{12}/c_{12}|^{2}+1} $ and $z_{12} = \frac{|b_{12}/d_{12}|^{2}}{|b_{12}/d_{12}|^{2}+1} $. Using the outlined prescription, the two-time correlation function $C_{ij}$ can be expressed as:
\begin{equation}
C_{ij} = 1-2 \ z_{ij} + 2 x_{ij}  \ (y_{ij}+ z_{ij}-1),
\label{E24} 
\end{equation}
with $0 \leq x_{ij}, y_{ij}, z_{ij} \leq 1$,and  $i<j \ \{i,j = 1,2,3\}$, where $x_{ij}  = \frac{r^{2}}{1+r^{2}}$, $y_{ij}= \frac{|a_{ij}/c_{ij}|^{2}}{|a_{ij}/c_{ij}|^{2}+1} $ and $z_{ij} = \frac{|b_{ij}/d_{ij}|^{2}}{|b_{ij}/d_{ij}|^{2}+1} $ (see \textcolor{blue}{Appendix \ref{AppB}} for more details). Here $x_{ij}$'s depend on the state under evolution, while $y_{ij}$'s and $z_{ij}$'s depend on the elements of the maps $f_{12}$, $f_{23}$ and $f_{12} \circ f_{23}$. It is interesting to note that $C_{ij}$'s have no initial state dependence for $(y_{ij}+ z_{ij}-1) = 0$ which is similar to the condition of stationarity defined in the context of unitary dynamics  in \cite{LGIreview2014}. 
\subsection{Exploration of violation of L\"{u}ders bound\label{IV-B}}
There is no clear prescription in the literature that which kind of qubit dynamics: (\textit{i}) unitary (linear  e.g closed system quantum dynamics),(\textit{ii}) non-unitary (linear e.g open system Lindbladian quantum dynamics) and (\textit{iii})non-unitary (non-linear e.g. measurement induced non-Hermitian quantum dynamics) can possibly satisfy or violate L\"{u}ders bound. It is evident that the violation of L\"{u}ders bound is impossible within unitary (linear) dynamics of a qubit (see \textcolor{blue}{Appendix \ref{AppC1}}) \cite{Anant,Sayan}. It is also established in a recent work \cite{Sayan} that qubit dynamics represented by unital maps (a subset of non-unitary (linear) dynamics) can never exceed L\"{u}ders bound over the full parameter space. On the contrary, other recent works \cite{Anant,Usha, Yogesh} have explored the violation of LG parameter $K_3$ beyond L\"{u}ders bound in non-Hermitian dynamics which is a subset of non-unitary (non-linear) dynamics. In this study, we are providing a mathematical constraint on the FLC map (\textit{although not exhaustive}) to satisfy L\"{u}ders bound irrespective of its linear (or non-linear) nature. Firstly a demonstrative example of non-unitary(non-linear) qubit dynamics satisfying L\"{u}ders bound case is shown which is induced by FLC maps. A general $2 \times 2$ non-unitary(non-linear) operator to evolve an initial qubit from a time instance $t_i$ to $t_j$, can be parametrized as,
\begin{equation}
\tilde{U}_{ij} =  \alpha_{ij} \mathbf{I} + \beta_{ij} \mathbf{\sigma_x}+ \gamma_{ij} \mathbf{\sigma_y}+ \zeta_{ij} \mathbf{\sigma_z} = \begin{pmatrix}
\alpha_{ij}+\zeta_{ij} & \beta_{ij}-i \gamma_{ij}
 \\ \beta_{ij}+i \gamma_{ij} & \alpha_{ij} - \zeta_{ij}
 \end{pmatrix}
 \label{E25}
\end{equation}
with $\{i,j = 1,2,3 \And i<j \}$, \ $\{\alpha_{ij},\ \beta_{ij},\ \gamma_{ij},\ \zeta_{ij} \in \mathbb{C}\}$ and $\{ \mathbf{I}, \mathbf{\sigma_{x/y/z}} \} $ being Pauli matrices. The FLC map corresponding to the $\tilde{U}_{ij}$ has the form:
\begin{equation}
f_{ij}(z) = \frac{(\alpha_{ij}+\zeta_{ij}) z + (\beta_{ij}-i \gamma_{ij}) }{(\beta_{ij}+i \gamma_{ij}) z + (\alpha_{ij} - \zeta_{ij})}
\label{E26}
\end{equation}
Considering $\boxed{\zeta_{ij} = 0; \ \gamma_{ij} = 0}$, the FLC maps $f_{12}$ and $f_{23}$ reduces to
\begin{equation}
f_{12}(z) = \frac{\alpha_{12}z + \beta_{12}}{\beta_{12} z + \alpha_{12} }, \ f_{23}(z) = \frac{\alpha_{23}z + \beta_{23}}{\beta_{23} z + \alpha_{23} } 
\label{E27}
\end{equation}
Further numerical calculations (see \textcolor{blue}{Appendix \ref{AppC2}}) and the plot (\textcolor{blue}{Fig. \ref{fig2}}) confirms that the above FLC map respects L{\"{u}}ders bound of $3/2$ for the full parameter space.
\par It is worth noting 
from eq.(\ref{E24}) that, if certain ratio are constrained in the following manner:
\begin{equation}
|a_{ij}/c_{ij}|= |d_{ij}/b_{ij}|, \ \text{such that} \ y_{ij}+ z_{ij} = 1 
\label{E28} 
\end{equation} 
then the upper bound of the LG parameter,  
\begin{equation}
K_3 = 1 - 2 z_{12} - 2 z_{23} + 2 z_{13}.
\label{E29} 
\end{equation}
always remains below the L{\"{u}}ders bound (refer to \textcolor{blue}{Appendix \ref{AppC2}} for more detailed calculation). The maximization of the L{\"{u}}ders bound for the $K_3$ parameter is independent of the initial qubit states in the Bloch sphere. See \textcolor{blue}{Table \ref{table:1}} for examples of such FLC maps.

\begin{table}[h!]
\centering
\begin{tabular}{ |p{2.1cm}|p{2cm}| }
\hline
\multicolumn{2}{|c|}{FLC Maps satisfying ratio constraint} \\ 
\hline
\vspace{0.02cm}\hspace{0.5cm}$\Large{f_{ij}(z)}$  & \vspace{0.02cm}\hspace{0.5cm}$\Large{f_{ij}(z)}$
\\ [3ex]
\hline 
\vspace{0.02cm}(i) $\dfrac{a z \pm b}{b z \pm a}$  &  \vspace{0.02cm}iii) $\dfrac{a z \pm b}{b^{*} z \pm a^{*}}$ \\ [4ex]
\vspace{0.02cm}(ii) $\dfrac{a z + b}{-b z \pm a}$  &  \vspace{0.02cm}iv)$\dfrac{a z + b}{-b^{*} z \pm a^{*}}$\\ [4ex]
\hline
\end{tabular}
\caption{FLC Maps satisfying ratio constraint and L{\"{u}}ders bound}
\label{table:1}
\end{table}

%(a) Trivial class involving linearity of quantum dynamics:  Given a state vector $\vert \psi \rangle = \alpha  \ \vert 0 \rangle + \beta \  \vert 1 \rangle$, linearity of the quantum dynamics implies existence of an operator $\hat{O}$: equation $\hat{O} \ \vert \psi \rangle = \alpha \ \hat{O}  \ \vert 0 \rangle + \beta \ \hat{O} \ \vert 1 \rangle$ is satisfied \cite{}. 
%(b)  non-trivial class which breaks linearity of quantum dynamics and yet respects L\"{u}ders bound \\
%(c)  non-trivial class which breaks both linearity of quantum dynamics and L\"{u}ders bound.
%%%%%%%%%%%%%%%%%%%%%%%%%%%%%%%%%%%%%%%%%%%%%%%%%%%%%%%%%%%%%%%%%%%%%%%%%%%%%%%%%%%%%%%%%%%%%%%%%%%%%%%%%%%%%%%%%%%%%%%%%%%%%%%%%%%%%%%%%%%%%%%%%%%%%%%%%%%%%%%%%%%%%%%%%%%%%%%%%%%%%%%%%%%%%%%%
\section{NSIT and AoT conditions\label{V}}
In this section, we relook at the considerations surrounding the interpretation of $K_3 > 1$ as an indicator of non-classical (quantum and beyond) dynamics. In recent years, various endeavors have been undertaken to address the noninvasive measurability loophole \cite{AKPan, Majidy} and the clumsiness loophole \cite{Huffman}. However, our focus here is exclusively on the statistical version of noninvasive measurability (NSIT) and arrow of time (AoT) conditions. It is crucial to recall that the simultaneous nonviolations of NSIT and AoT conditions ensure the existence of a global joint probability (\textcolor{blue}{see Appendix \ref{AppA}}) distribution \cite{Clemente}, implying macroscopic realism. In the context of unitary dynamics, NSIT conditions are typically violated, while all AoT conditions are satisfied. Considering the dynamical process induced on a qubit by fractional linear conformal maps $f_{12}$, $f_{23}$, and $f_{13}$, explicit calculations (refer to \textcolor{blue}{Appendix \ref{AppA}}) reveal the following: (a) all two-time AoT conditions of the form $\text{AoT}_{{i(j)}} : P(m_i) = \sum_{m_j = \pm 1} P(m_i, m_j)$ are satisfied (where $\{i,j = 1,2,3 \And i<j\}$; (b) three-time AoT conditions of types $\text{AoT}_{12(3)} : P(m_1,m_2) = \sum_{m_3 = \pm 1} P(m_1, m_2, m_3)$ and $\text{AoT}_{1(23)} : P(m_1) \equiv \sum_{m_2,m_3 = \pm 1} P(m_1, m_2, m_3)$ are always satisfied; and (c) all NSIT conditions of types $\text{NSIT}_{(i)j} : P(m_j) = \sum_{m_i = \pm 1} P(m_i,m_j), \quad \quad i < j$, and $\text{NSIT}_{1(2)3} : P(m_1,m_3) \equiv \sum_{m_2 = \pm 1} P(m_1,m_2,m_3)$ and $\text{NSIT}_{(1)23} : P(m_2,m_3) \equiv \sum_{m_1 = \pm 1} P(m_1,m_2,m_3)$ are generally not satisfied (here $P(m_1,m_2,m_3)$ denotes the global joint probability with $m_1,m_2,m_3$ being the outcomes of dichotomic observable at time instances $t_1, t_2$ and $t_3$ respectively. In general, the dynamics induced by the presented fractional linear conformal maps are found to be inconsistent with Macroscopic Realism (MR).
\par In the rest of this section we present an example of FLC map (having non-linear and non-unitary) action on the state space) for which $\text{NSIT}$ of the type $\text{NSIT}_{1(2)3}$ is violated throughout the parameter space (except for a one parameter family) of map elements . We again take the FLC maps to be following,
\begin{equation}
f_{12}(z) = \frac{\alpha_1 \ z + \beta_1}{\beta_1 \ z + \alpha_1}, \ \  \text{and} \ \  f_{23}(z) = \frac{\alpha_2 \ z + \beta_2}{\beta_2 \ z + \alpha_2}, \\
\label{E34} 
\end{equation}
\begin{equation*}
z_3 = f_{23}(f_{12}(z)) = \frac{(\alpha_1 \alpha_2 + \beta_1 \beta_2)z + (\beta_1 \alpha_2 + \alpha_1 \beta_2)}{(\beta_1 \alpha_2 + \alpha_1 \beta_2)z + (\alpha_1 \alpha_2 + \beta_1 \beta_2)} 
\end{equation*}
\begin{equation}
= \frac{a_{13}z_1 + b_{13}}{c_{13} z_1 + d_{13}}
\label{E35}
\end{equation}
A thorough calculation of joint probabilities (where outcome of the dichotomic observable is only $+1$ for all time instances (\textcolor{blue}{see Appendix \ref{AppD} for other outcomes}) using the above FLC maps yields (see \textcolor{blue}{Appendix \ref{AppA}}),
\begin{equation}
P(+^1, +^3) = \frac{r^2}{1+r^2} \Big( \frac{\vert \lambda_1 \lambda_2 + 1 \vert^2}{\vert \lambda_1 \lambda_2 + 1 \vert^2 + \vert \lambda_1 + \lambda_2 \vert^2} \big)
\label{E36}
\end{equation}
\begin{equation}
P(+,+,+) + P(+,-, +) = \frac{r^2}{1+r^2} \frac{\vert \lambda_1 \lambda_2 
\vert^2+ 1 }{(\vert \lambda_1 \vert^2  + 1 ) ( \vert \lambda_2 \vert^2 +1)} 
\label{E37}
\end{equation}
where $\lambda_1 = \frac{\beta_1}{\alpha_1}$ and $\lambda_2 = \frac{\beta_2}{\alpha_2}$.\\ \\
For $\text{NSIT}_{1(2)3}: P(+^1, +^3) = P(+,+,+) + P(+,-, +)$ to be obeyed both the eq.(\ref{E36}) and eq.(\ref{E37}) must match, yielding the following conditions,
\begin{equation}
\text{Re}(\lambda_1 \lambda_2) = 0 \quad \quad \text{and} \quad \quad \text{Re}(\lambda_1) \text{Re}(\lambda_2) = 0
\label{} \nonumber
\end{equation}
which in turn has two solutions,
\begin{equation}
\text{Condition \ 1:} \quad \quad \text{Re}(\lambda_1) = \text{Im}(\lambda_2) = 0
\label{E38}
\end{equation}
\begin{equation}
\text{Condition \ 2:} \quad \quad \text{Re}(\lambda_2) = \text{Im}(\lambda_1) = 0
\label{E39}
\end{equation}
This shows that the the $\text{NSIT}_{1(2)3}$ condition is only satisfied for i)\ $\vert \delta / \gamma \vert = 0$ ( corresponding to eq.(\ref{E38})), ii) \  $\vert \beta / \alpha \vert = 0$ ( corresponding to eq.(\ref{E39}))(\textcolor{blue}{see figure \ref{fig2}(a)}). The above found parameter space also ensures that LG parameter $K_3$ is bounded by $1$ confirming classical behaviour of the dynamics. In general, for most of the parameter space of the FLC maps, optimal $K_3$ value is in the non-classical regime whenever NSIT conditions are violated.

\section{Conclusion\label{VI}}
In this study, We have established that  the fractional linear conformal maps encompass a large variety of conceivable quantum dynamics interrupted (uninterrupted) by quantum measurements. We explore temporal correlations, quantified by the Leggett-Garg parameter $K_3$, in a two-level system subjected to the dynamics induced by fractional linear conformal maps, specifically denoted by $f_{12}$, $f_{23}$, and $f_{13}$. Our findings demonstrate that, when certain ratio constraints among the elements of these maps are met, the $K_3$ parameter remains within the confines of the Lüders bound, never surpassing the $3/2$ bound.
%In this paper, we have investigated the temporal correlations quantified in terms of LG parameter $K_3$ for a two-level system whose dynamics is induced by fractional linear conformal maps $f_{12}$, $f_{23}$ and $f_{13}$. We showed that if one satisfies the \textit{ratio constraints} among the elements of the maps, LG parameter $K_3$ never exceeds the L{\"{u}}ders bound of $3/2$.
\par Additionally, we presented illustrative examples of the aforementioned maps in the form of a table, including the one falling under the category of \textit{unitary dynamics.} Our investigation categorizes the classes of fractional linear conformal maps into three major distinct groups: (\textit{i}) the action on the qubit state space is linear, and the Lüders bound is respected; (\textit{ii})  the action on the state space is non-linear yet respects the Lüders bound; and (\textit{iii}) the action on the state space is non-linear  and  violates the Lüders bound if specific \textit{ratio} constraints are not satisfied.

\section{Acknowledgement} 
S.P. offers his gratitude to the Council of Scientific and
Industrial Research (CSIR), Govt. of India for financial
support.

\pagebreak

\color{black}{

\appendix

\section{Global joint probabilities \label{AppA}}

\begin{widetext}
Using the maps $f
_{12}$ ,  $f_{23}$  and  $f_{13}$ (given in main text eq.(\ref{E08})), we calculate the global three time joint probabilities as following,
\begin{equation*}
P(+,+,+)  = {|\langle \uparrow \vert\psi(t_3) \rangle_{\ket{\uparrow}}|^2|\langle \uparrow \vert\psi(t_2) \rangle_{\ket{\uparrow}}|^2|\langle \uparrow \vert\psi(t_1) \rangle|^2} =  \frac{|a_{23}/c_{23}|^{2}}{|a_{23}/c_{23}|^{2}+1} \times 
\frac{|a_{12}/c_{12}|^{2}}{|a_{12}/c_{12}|^{2}+1} \times \frac{r^{2}}{1+r^{2}} 
\end{equation*}
\begin{equation*}
P(+,+,-)  = {|\langle \downarrow \vert\psi(t_3) \rangle_{\ket{\uparrow}}|^2|\langle \uparrow \vert\psi(t_2) \rangle_{\ket{\uparrow}}|^2|\langle \uparrow \vert\psi(t_1) \rangle|^2} = \frac{1}{|a_{23}/c_{23}|^{2}+1} \times 
\frac{|a_{12}/c_{12}|^{2}}{|a_{12}/c_{12}|^{2}+1} \times \frac{r^{2}}{1+r^{2}} 
\end{equation*}
\begin{equation*}
P(+,-,+)  = {|\langle \uparrow \vert\psi(t_3) \rangle_{\ket{\downarrow}}|^2|\langle \downarrow \vert\psi(t_2) \rangle_{\ket{\uparrow}}|^2|\langle \uparrow \vert\psi(t_1) \rangle|^2} =  \frac{|b_{23}/d_{23}|^{2}}{|b_{23}/d_{23}|^{2}+1} \times 
\frac{1}{|a_{12}/c_{12}|^{2}+1} \times \frac{r^{2}}{1+r^{2}} 
\end{equation*}
\begin{equation*}
P(+,-,-)  = {|\langle \downarrow \vert\psi(t_3) \rangle_{\ket{\downarrow}}|^2|\langle \downarrow \vert\psi(t_2) \rangle_{\ket{\uparrow}}|^2|\langle \uparrow \vert\psi(t_1) \rangle|^2} =  \frac{1}{|b_{23}/d_{23}|^{2}+1} \times 
\frac{1}{|a_{12}/c_{12}|^{2}+1} \times \frac{r^{2}}{1+r^{2}} 
\end{equation*}
\begin{equation*}
P(-,+,+)  = {|\langle \uparrow \vert\psi(t_3) \rangle_{\ket{\uparrow}}|^2|\langle \uparrow \vert\psi(t_2) \rangle_{\ket{\downarrow}}|^2|\langle \downarrow \vert\psi(t_1) \rangle|^2} =  \frac{|a_{23}/c_{23}|^{2}}{|a_{23}/c_{23}|^{2}+1} \times 
\frac{|b_{12}/d_{12}|^{2}}{|b_{12}/d_{12}|^{2}+1} \times \frac{1}{1+r^{2}} 
\end{equation*}
\begin{equation*}
P(-,+,-)  = {|\langle \downarrow \vert\psi(t_3) \rangle_{\ket{\uparrow}}|^2|\langle \uparrow \vert\psi(t_2) \rangle_{\ket{\downarrow}}|^2|\langle \downarrow \vert\psi(t_1) \rangle|^2} = \frac{1}{|a_{23}/c_{23}|^{2}+1} \times 
\frac{|b_{12}/d_{12}|^{2}}{|b_{12}/d_{12}|^{2}+1} \times \frac{1}{1+r^{2}} 
\end{equation*}
\begin{equation*}
P(-,-,+)  = {|\langle \uparrow \vert\psi(t_3) \rangle_{\ket{\downarrow}}|^2|\langle \downarrow \vert\psi(t_2) \rangle_{\ket{\downarrow}}|^2|\langle \downarrow \vert\psi(t_1) \rangle|^2} =  \frac{|b_{23}/d_{23}|^{2}}{|b_{23}/d_{23}|^{2}+1} \times 
\frac{1}{|b_{12}/d_{12}|^{2}+1} \times \frac{1}{1+r^{2}} 
\end{equation*}
\begin{equation}
P(-,-,-)  = {|\langle \downarrow\vert\psi(t_3) \rangle_{\ket{\downarrow}}|^2|\langle \downarrow\vert\psi(t_2) \rangle_{\ket{\downarrow}}|^2|\langle \downarrow \vert\psi(t_1) \rangle|^2} =  \frac{1}{|b_{23}/d_{23}|^{2}+1} \times 
\frac{1}{|b_{12}/d_{12}|^{2}+1} \times \frac{1}{1+r^{2}} 
\label{A1}
\end{equation}
Similarly we procced to calculate the two time joint ($t_1,t_2)$ probabilities:
\begin{equation*}
P(+^1,+^2) = |\langle \uparrow\vert\psi(t_2) \rangle_{\ket{\uparrow}}|^2|\langle \uparrow \vert\psi(t_1) \rangle|^2 = 
\frac{|a_{12}/c_{12}|^{2}}{|a_{12}/c_{12}|^{2}+1} \times \frac{r^{2}}{1+r^{2}} 
\end{equation*}
\begin{equation*}
P(+^1,-^2) = |\langle \downarrow \vert\psi(t_2) \rangle_{\ket{\uparrow}}|^2|\langle \uparrow \vert\psi(t_1) \rangle|^2 = 
\frac{1}{|a_{12}/c_{12}|^{2}+1} \times \frac{r^{2}}{1+r^{2}} 
\end{equation*}
\begin{equation*}
P(-^1,+^2) = |\langle \uparrow\vert\psi(t_2) \rangle_{\ket{\downarrow}}|^2|\langle \downarrow \vert\psi(t_1) \rangle|^2 = 
\frac{|b_{12}/d_{12}|^{2}}{|b_{12}/d_{12}|^{2}+1} \times \frac{1}{1+r^{2}} 
\end{equation*}
\begin{equation}
P(-^1,-^2) = |\langle\downarrow\vert\psi(t_2) \rangle_{\ket{\downarrow}}|^2|\langle \downarrow \vert\psi(t_1) \rangle|^2 = 
\frac{1}{|b_{12}/d_{12}|^{2}+1} \times \frac{1}{1+r^{2}} 
\label{A2}
\end{equation}
where $P(\pm^{i}, \pm^{j}) $ denotes that the measurement is done at time instances $t_i$ first and then $t_j$. Next we have the two time joint ($t_1,t_3)$ probabilities following:
\begin{equation*}
P(+^1,+^3) = |\langle \uparrow\vert\psi(t_3) \rangle_{\ket{\uparrow}}|^2|\langle \uparrow \vert\psi(t_1) \rangle|^2 = 
\frac{|a_{13}/c_{13}|^{2}}{|a_{13}/c_{13}|^{2}+1} \times \frac{r^{2}}{1+r^{2}} 
\end{equation*}
\begin{equation*}
P(+^1,-^3) = |\langle \downarrow \vert\psi(t_3) \rangle_{\ket{\uparrow}}|^2|\langle +_z \vert\psi(t_1) \rangle|^2 = 
\frac{1}{|a_{13}/c_{13}|^{2}+1} \times \frac{r^{2}}{1+r^{2}} 
\end{equation*}
\begin{equation*}
P(-^1,+^3) = |\langle \uparrow \vert\psi(t_3) \rangle_{\ket{\downarrow}}|^2|\langle \downarrow \vert\psi(t_1) \rangle|^2 = 
\frac{|b_{13}/d_{13}|^{2}}{|b_{13}/d_{13}|^{2}+1} \times \frac{1}{1+r^{2}} 
\end{equation*}
\begin{equation}
P(-^1,-^3) = |\langle \downarrow \vert\psi(t_3) \rangle_{\ket{\downarrow}}|^2|\langle \downarrow \vert\psi(t_1) \rangle|^2 = 
\frac{1}{|b_{13}/d_{13}|^{2}+1} \times \frac{1}{1+r^{2}} 
\label{A3}
\end{equation}
Next we have the two time joint ($t_2,t_3)$ probabilities following:
\begin{equation*}
P(+^2,+^3) = |\langle \uparrow \vert\psi(t_3) \rangle_{\ket{\uparrow}}|^2|\langle \uparrow \vert\psi(t_2) \rangle|^2 = 
\frac{|a_{23}/c_{23}|^{2}}{|a_{23}/c_{23}|^{2}+1} \times \frac{|(a_{12} z_1 + b_{12})/(c_{12} z_1 + d_{12})|^2}{1+|(a_{12} z_1 + b_{12})/(c_{12} z_1 + d_{12})|^2} 
\end{equation*}
\begin{equation*}
P(+^2,-^3) = |\langle \downarrow \vert\psi(t_3) \rangle_{\ket{\uparrow}}|^2|\langle \uparrow \vert\psi(t_2) \rangle|^2 = 
\frac{1}{|a_{23}/c_{23}|^{2}+1} \times \frac{|(a_{12} z_1 + b_{12})/(c_{12} z_1 + d_{12})|^2}{1+|(a_{12} z_1 + b_{12})/(c_{12} z_1 + d_{12})|^2} 
\end{equation*}
\begin{equation*}
P(-^2,+^3) = |\langle \uparrow \vert\psi(t_3) \rangle_{\ket{\downarrow}}|^2|\langle \downarrow \vert\psi(t_2) \rangle|^2 = 
\frac{|b_{23}/d_{23}|^{2}}{|b_{23}/d_{23}|^{2}+1} \times \frac{1}{1+|(a_{12} z_1 + b_{12})/(c_{12} z_1 + d_{12})|^2} 
\end{equation*}
\begin{equation}
P(-^2,-^3) = |\langle \downarrow \vert\psi(t_3) \rangle_{\ket{\downarrow}}|^2|\langle \downarrow \vert\psi(t_2) \rangle|^2 = 
\frac{1}{|b_{23}/d_{23}|^{2}+1} \times \frac{1}{1+|(a_{12} z_1 + b_{12})/(c_{12} z_1 + d_{12})|^2} 
\label{A4}
\end{equation}
where $z_1 = r e^{i \phi}$ which corresponds to the initial unmeasured state $\vert \psi(t_1) \rangle = \vert \psi^{(0)} \rangle = \frac{1}{\sqrt{\vert z_1 \vert^2 + 1}} (z_1 , 1)$. \\ \\
The one time ($t_i$) probabilities:
\begin{equation*}
P(+^1) = |\langle \uparrow \vert\psi(t_1) \rangle|^2 =  \frac{r^{2}}{1+r^{2}} , \quad \quad P(-^1) = |\langle \downarrow \vert\psi(t_1) \rangle|^2 = \frac{1}{1+r^{2}} 
\end{equation*}
\begin{equation*}
P(+^2) = |\langle \uparrow \vert\psi(t_2) \rangle|^2 =   \frac{|(a_{12} z_1 + b_{12})/(c_{12} z_1 + d_{12})|^2}{1+|(a_{12} z_1 + b_{12})/(c_{12} z_1 + d_{12})|^2} , \quad \quad P(-^2) = |\langle \downarrow \vert\psi(t_2) \rangle|^2 = \frac{1}{1+|(a_{12} z_1 + b_{12})/(c_{12} z_1 + d_{12})|^2} 
\end{equation*}
\begin{equation}
P(+^3) = |\langle \uparrow \vert\psi(t_3) \rangle|^2 =   \frac{|(a_{13} z_1 + b_{13})/(c_{13} z_1 + d_{13})|^2}{1+|(a_{13} z_1 + b_{13})/(c_{13} z_1 + d_{13})|^2}  , \quad \quad P(-^3) = |\langle \downarrow \vert\psi(t_3) \rangle|^2 =  \frac{1}{1+|(a_{13} z_1 + b_{13})/(c_{13} z_1 + d_{13})|^2} 
\label{A5}
\end{equation}
\\ \\
\end{widetext}

\section{Sample calculation of $C_{ij}$: \label{AppB}}

Following the same argument given in the main text, $C_{23}$ can be expressed as:
\begin{equation}
C_{23} = 1-2 z_{23} + 2 x_{23} (y_{23}+z_{23}-1),
\label{B01}
\end{equation}
where $y_{23}= \frac{|a_{23}/c_{23}|^{2}}{|a_{23}/c_{23}|^{2}+1} $ and $z_{23} = \frac{|b_{23}/d_{23}|^{2}}{|b_{23}/d_{23}|^{2}+1} $.\\ \\
Likewise the joint probabilities for the correlation $C_{13}$ are as follows:

\begin{equation}
 P_{13}(+,+)= \frac{r^{2}}{1+r^{2}} \times y_{13}
 \label{B02}
\end{equation}

\begin{equation}
 P_{13}(+,-)=  \frac{r^{2}}{1+r^{2}} \times   (1-y_{13}) 
 \label{B03}
\end{equation}

\begin{equation}
 P_{13}(-,+)= \frac{1}{1+r^{2}} \times  z_{13}  
 \label{B04}
\end{equation}

\begin{equation}
 P_{13}(-,-)= \frac{1}{1+r^{2}} \times (1-z_{13})
\label{B05}
\end{equation}

\begin{equation}
C_{13} = 1-2 z_{13} + 2 x_{13} (y_{13}+z_{13}-1),
\label{B06}
\end{equation}

where $y_{13}$ and $z_{13}$ are given by:

\begin{equation}
x_{13} = \frac{r^2}{1+r^2}, \quad \quad y_{13} = \frac{|a_{13}/c_{13}|^{2}}{|a_{13}/c_{13}|^{2}+1}
\label{B07}
\end{equation}

\begin{equation}
z_{13} = \frac{|b_{13}/d_{13}|^{2}}{|b_{13}/d_{13}|^{2}+1} \\
\label{B08}
\end{equation}

\section{Derivation of L{\"{u}}ders bound for (i)\ unitary (linear) case of qubit, (ii)\ general non-unitary (non-linear) case of FLC Maps \label{AppC}}

\subsection{Unitary Case\label{AppC1}}
For unitary FLC maps (given in main text eq.(\ref{E08})) the elements of $f_{12}$, $f_{23}$ and $f_{13}$ are related by,
\begin{equation}
d_{ij} = a^{*}_{ij}, \quad c_{ij} = -b^{*}_{ij}
\label{C1}
\end{equation}
together with $$ |a_{ij}|^2 + |b_{ij}|^2 = 1 \ \text{for} \ i < j ,\{i,j = 1,2,3\}$$
Hence the expressions of $C_{ij}$'s simplify to,
\begin{equation*}
C_{12} = 1 - 2 |b_{12} |^2, \  C_{23} = 1 - 2 |b_{23} |^2 
\end{equation*}
\begin{equation*}
C_{13} = 1 - 2 |b_{13} |^2
\end{equation*}
\begin{equation}
= 1- 2 \Big( |a_{12}|^2 |b_{23}|^2 + |b_{12}|^2 |a_{23}|^2 + 2 \text{Re}(a_{12} a_{23} b_{12} b^{*}_{23}) \Big)
\label{C2}
\end{equation}
Taking the quantities $a_{12} = \cos(\theta_1) e^{i \gamma_1},\ b_{12} = \sin(\theta_1) e^{i \gamma_2}, \ a_{23} = \cos(\theta_2) e^{i \gamma_3},\ b_{23} = \sin(\theta_2) e^{i \gamma_4}$, the expression of $K_3$ becomes
\begin{equation*}
K_3 = \cos 2\theta_1 + \cos 2\theta_2 - \cos 2\theta_1 \cos 2\theta_2 
\end{equation*}
\begin{equation}
- \sin 2\theta_1 \sin 2\theta_2 \cos 2\gamma
\label{C3}
\end{equation}
where $\gamma = \gamma_1 + \gamma_2 + \gamma_3 - \gamma_4$.
\par Eq.(\ref{C3}) ensures that $K_3$ is upper bounded by L{\"{u}}ders bound. 

\subsection{General case \label{AppC2}}
Following the ratio constraints (given in main text eq.(\ref{E28})) the expression of LG parameter becomes $K_3 = 1- 2 z_{12} - 2 z_{23} + 2 z_{13}$. Assuming the following quantities: \  $b_{12}/d_{12} = r_1 e^{i \theta_1},\ b_{23}/d_{23} = r_2 e^{i \theta_2},\ a_{23}/b_{23} = r_3 e^{i \theta_3},\ c_{23}/d_{23} = r_4 e^{i \theta_4}$, we obtain
\begin{equation*}
z_{12} = \frac{r_1^{2}}{1+ r_1^{2}}, \quad \quad \quad z_{23} = \frac{r_2^{2}}{1+ r_2^{2}}
\end{equation*}
\begin{equation}
z_{13} = \frac{A}{A+B}
\label{C4}
\end{equation}
with $A = r_2^{2} \Big(r_1^{2} r_3^{2} + 1 + 2 r_1 r_3 \cos(\theta_1 + \theta_3 )  \Big)$ and $B = r_1^{2}r_4^{2} + 1 + 2 r_1 r_4 \cos(\theta_1 +\theta_4)$. Employing the ratio constraint $\vert \frac{a_{12}}{c_{12}} \vert = \vert \frac{d_{12}}{b_{12}} \vert $ and $\vert \frac{a_{23}}{c_{23}} \vert = \vert \frac{d_{23}}{b_{23}} \vert $, it can be established  \begin{equation}
\boxed{r_4^2 = r_3^2 r_2^4}    
\label{C5}
\end{equation}. 
From the ratio constraint $\vert \frac{a_{13}}{c_{13}} \vert = \vert \frac{d_{13}}{b_{13}} \vert $ we also find another relation ,
\begin{equation}
r_2^2 \Big( \frac{r_1^2 + r_3^2 + 2 r_1 r_3 \cos\theta_3}{r_1^2 + r_4^2 + 2 r_1 r_4 \cos\theta_4} \Big)= \big(\frac{r_1^2 r_4^2 + 1 + 2 r_1 r_4 \cos\theta_4}{r_1^2 r_3^2 + 1 + 2 r_1 r_3 \cos\theta_3}
\label{C6}
\end{equation}
Likewise $\theta_1$ can be set to \textbf{zero} (without loss of generality). Hence $K_3$ parameter becomes a function of \textit{four} unknown parameters \{ $r_1,\ r_2,\ \theta_3, \ \theta_4$ \}. It is checked numerically that, $K_3$ respects L{\"{u}}ders bound for the full parameter space of \{ $r_1,\ r_2,\ \theta_3, \ \theta_4$ \} .
\section{NSIT conditions \label{AppD}}
It is evident from (Appendix \ref{A1}-\ref{A4})  $\text{NSIT}_{1(2)3}$ is violated for all spin projections i.e.:
\begin{equation}
P(+^1,+^3) \neq P(+,+,+)+P(+,-,+)
\label{D1}
\end{equation}
\begin{equation}
P(+^1,-^3) \neq P(+,+,-)+P(+,-,-)
\label{D2}
\end{equation}
\begin{equation}
P(-^1,+^3) \neq P(-,+,+)+P(-,-,+)
\label{D3}
\end{equation}
\begin{equation}
P(-^1,-^3) \neq P(-,+,-)+P(-,-,-)
\label{D4}
\end{equation}
Likewise all other NSIT conditions of type $\text{NSIT}_{(1)2}$, $\text{NSIT}_{(2)3}$, $\text{NSIT}_{(1)3}$ and $\text{NSIT}_{(1)23}$ are also violated for arbitrary parameter values.\\

\end{document}